# Rapid and reliable thickness identification of two-dimensional nanosheets using optical microscopy**

*Hai Li,[1,†] Jumiati Wu,[1,†] Xiao Huang,[1] Gang Lu,[1] Jian Yang,[1] Xin Lu,[2] Qihua Xiong,[2] Hua Zhang[1,*]*

[1]School of Materials Science and Engineering, Nanyang Technological University, 50 Nanyang Avenue, Singapore 639798, Singapore

[2]Division of Physics and Applied Physics, School of Physical and Mathematical Sciences, Nanyang Technological University, 21 Nanyang Link, Singapore 637371, Singapore

*To whom correspondence should be addressed.

Phone: +65-6790-5175. Fax: +65-6790-9081

E-mail: hzhang@ntu.edu.sg

Website: http://www.ntu.edu.sg/home/hzhang/

[†]These authors contributed equally to this work.

**ABSTRACT.** The physical and electronic properties of ultrathin two-dimensional (2D) layered nanomaterials are highly related to their thickness. Therefore, the rapid and accurate identification of single- and few- to multi-layer nanosheets is essential to their fundamental study and practical applications. Here, a universal optical method has been developed for simple, rapid and reliable identification of single- to quindecuple-layer (1L-15L) 2D nanosheets, including graphene, $MoS_2$, $WSe_2$ and $TaS_2$, on Si substrates coated with 90 nm or 300 nm $SiO_2$. The optical contrast differences between the substrates and 2D nanosheets with different layer numbers were collected and tabulated, serving as a standard reference, from which the layer number of a given nanosheet can be readily and reliably determined without using complex calculation nor expensive instrument. Our general optical identification method will facilitate the thickness-dependent study of various 2D nanomaterials, and expedite their research toward practical applications.

Two-dimensional (2D) layered nanomaterials, such as graphene and transition metal dichalcogenides (TMDs, e.g. $MoS_2$, $MoSe_2$, $WS_2$, $WSe_2$, $NbSe_2$, $TiS_2$, and $TaS_2$),[1-10] have attracted much attention in recent years due to their novel optical, electronic, mechanical and magnetic properties in contrast to their bulk crystals. Currently, mechanical exfoliation is still one of the most efficient ways to obtain high-quality, atomically thin nanosheets of 2D layered nanomaterials.[1, 3-7] However, this technique produces not only single- and few- to multi-layer nanosheets, but also a large quantity of thicker flakes. It is well known that the physical and electronic properties of 2D nanomaterials are highly related to their thickness.[3-6, 11-17] For example, while single-layer (1L) graphene is a zero–band gap semimetal, double-layer (2L) graphene is semiconducting with tunable band gap, leading to much higher on/off current ratios in field-effect transistors (FETs).[12, 18] Few-layer graphene also shows different energy band structures from 1L graphene, and exhibits some favorable optical and electronic properties for practical applications.[12-13, 18] Similar to graphene, TMD nanosheets also show the thickness-dependent band structures. For example, the 1L and 2L $MoS_2$ nanosheets with band gap of 1.82 eV and 1.65 eV, respectively, are attractive for green light detection, while triple-layer (3L) $MoS_2$ nanosheet with a band gap of 1.32 eV is more sensitive to red light.[16] On the other hand, multilayer $MoS_2$ nanosheets are attractive for fabrication of flexible transparent devices, due to their ease of fabrication, good mechanical and electronic stability, and ability to provide high current drive in the devices.[14, 19] Therefore, the rapid determination of location and layer number of mechanically exfoliated single- and few- to multi-layer nanosheets among copious thick flakes over a centimeter-/millimeter-size area is the first priority in their fundamental research and practical applications.

To date, many methods have been developed to identify the thickness of 2D nanosheets, such as atomic force microscopy (AFM), Raman spectroscopy and optical microscopy (OM). Although AFM is commonly used to measure the thickness of 2D nanosheets, it is time-consuming and not suitable for rapid measurement over large area. In addition, AFM measurement might be affected by the absorbed water layer under 2D nanosheets or instrumental offset.[20-21] As a result, the thickness of single-layer graphene measured by AFM varied from 0.4 to 0.9 nm.[18, 21-22] Raman spectroscopy is a quick characterization method to identify single- to few-layer 2D nanosheets.[12, 17, 23-25] However, the difference between double- and few-layer graphene or TMDs nanosheets in Raman spectra is insufficient to accurately distinguish them.[4, 23-24] Although low-frequency Raman spectroscopy (< 50 cm$^{-1}$) has been used to reliably determine the layer number of graphene, MoS$_2$ or WSe$_2$,[12, 17, 25] it requires expensive and nonstandard equipment. On the contrary, OM is a simple, efficient and nondestructive technique that enables rapid characterization of 2D nanosheets over large area.[4, 13, 20, 26-35] The OM method mainly relies on the optical contrast between a 2D nanosheet and the substrate for fast and unambiguous identification. To improve such contrast, several methods have been developed, including the use of narrow band illumination,[28, 36] selection of optimal substrate,[26, 30, 36] collection of reflection spectra,[20] measurement of total color difference[30] or ratio of color difference[29] *etc*. Unfortunately, these methods either involve special experimental setup or time-consuming image processing, and more importantly, they are not generalizable for identification of various kinds of nanosheets.

Here, we demonstrate a simple, rapid and reliable method to identify 2D nanosheets (*e.g.* graphene, MoS$_2$, WSe$_2$ and TaS$_2$) from single- to quinduceple-layer (1L-15L) without using expensive instrument nor complex calculation. The contrast difference between the 2D nanosheet and substrate can be simply obtained from the brightness profile of their color images or grayscale images of R, G or B channel. The obtained values of contrast difference for nanosheets with different layer numbers can be plotted as a standard chart, based on which the layer number of a given nanosheet can be rapidly and accurately determined on Si substrate coated with 90 nm or 300 nm $SiO_2$, referred to 90 nm or 300 nm $SiO_2$/Si, respectively.

**RESULTS AND DISCUSSION**

**Description of the optical identification method**

The key to the reliable and accurate optical identification of a 2D nanosheet is to correlate its layer number with its optical contrast with respect to the substrate. In our method, the optical contrast of a nanosheet (defined as $C$) and substrate (defined as $Cs$) were directly measured from its color optical image by using a free software (ImageJ). The contrast difference (defined as $C_D$) is obtained by subtracting $C$ with $Cs$ (Equation 1). Similarly, for the grayscale image (from R, G or B channel), the contrast difference between the nanosheet and substrate ($C_{DR}$, $C_{DG}$ or $C_{DB}$) is calculated by subtracting the contrast of the nanosheet ($C_R$, $C_G$ or $C_B$) with that of the substrate ($C_{SR}$, $C_{SG}$ or $C_{SB}$) (Equation 2-4).

$$C_D = C - C_S \tag{1}$$

$$C_{DR} = C_R - C_{SR} \tag{2}$$

$$C_{DG} = C_G - C_{SG} \tag{3}$$

$$C_{DB} = C_B - C_{SB} \tag{4}$$

As a demonstration, Figure 1a shows the color optical image of a $MoS_2$ flake on 90 nm $SiO_2$/Si. Fig. 1b is the contrast profile of the dashed rectangle highlighted in Figure 1a generated by ImageJ. The contrast values (*C*) of the octuple-layer (8L) and hextuple-layer (6L) $MoS_2$ nanosheets are 162.3 and 118.6, respectively, while the contrast value of 90 nm $SiO_2$/Si ($C_S$) is 120.4. According to equation (1), the contrast difference between the 8L (or 6L) $MoS_2$ nanosheet and 90 nm $SiO_2$/Si substrate is calculated to be $C_D$ = 162.3 - 120.4 = 41.9 (or 118.6 -120.4 = -1.8).

**Optical identification of 1L-15L graphene nanosheets on 90 nm $SiO_2$/Si**

It has been reported that the color of a graphene nanosheet can be used to identify its thickness in combination with theoretical calculation.[30, 34] Theoretical calculation predicted that $SiO_2$ film with thickness of 90 or 300 nm is the optimal dielectric layer for optical identification of graphene.[13, 26, 35] Here, thickness identification of 1L-15L graphene nanosheets on 90 or 300 nm $SiO_2$/Si can be achieved by our simple, rapid and reliable method based on the measurement of optical contrast difference. Color optical images (Figure 2a-n), AFM measurement (Figure 2q) and Raman characterization (Supplementary Figure S1) were first used to locate exfoliated graphene nanosheets on 90 nm $SiO_2$/Si and determine their thicknesses. After that, optical contrast differences ($C_D$) between 1L-15L graphene nanosheets and 90 nm $SiO_2$/Si were measured from their color optical images taken at different exposure times by using ImageJ (Figure 2o). It can be seen that, for 1L-15L graphene nanosheets, the absolute value of $C_D$ increases with increasing exposure time from 20 to 140 ms, and decreases from 160 to 300 ms. At 200 ms, as compared to other exposure times, such as 80 ms, the $C_D$ values are mostly distinguishable among the 1L-15L nanosheets (Figure 2p and Supplementary Table S1), especially for those thicker than 10L. Therefore, a standard chart of $C_D$ values at 200 ms for different layer numbers was generated (Figure 2p and Supplementary Table S1), from which the thickness of a graphene nanosheet on 90 nm $SiO_2$/Si can be readily determined.

Similar to the $C_D$ values, the $C_{DR}$, $C_{DG}$ and $C_{DB}$ values of 1L-15L graphene nanosheets on 90 nm $SiO_2$/Si, measured from grayscale images of R, G and B channels, respectively, can also be used for the layer number identification (Supplementary Figure S2 and Table S1). Similarly, the $C_D$, $C_{DR}$ and $C_{DG}$ values of 1L-13L graphene nanosheets on 300 nm $SiO_2$/Si can also be determined in the same manner and thus used for layer number identification of graphene (Supplementary Figures S3-S4).

**Optical identification of 1L-15L $MoS_2$ nanosheets on 90 nm $SiO_2$/Si**

As for the 1L-15L $MoS_2$ nanosheets on 90 nm $SiO_2$/Si (Figure 3a-n), they also show the thickness-dependent contrast difference at various exposure times (Figure 3o-p), confirmed by AFM measurement and low-frequency Raman characterization (Figure 3q and Supplementary Figure S5). It is shown that at the exposure time of 80 ms, 5L-15L $MoS_2$ nanosheets are distinguishable based on $C_D$ values (Figure 3o-p and Supplementary Table S2). However, the $C_D$ values of 1L-4L $MoS_2$ nanosheets are less differentiable, and especially the $C_D$ values of 2L and 3L nanosheets are close, ~ -55 (Figure 3p and Figure 4 a-b). In order to effectively distinguish 1L-4L $MoS_2$ nanosheets on 90 nm $SiO_2$/Si, the optical contrast differences measured from their grayscale images from R, G and B channels ($C_{DR}$, $C_{DG}$ and $C_{DB}$) were used to determine their thicknesses (Figure 4c-i). As shown in Figure 4g-i, the $C_{DB}$ values of 1L and 2L $MoS_2$ nanosheets are negative (1L: - 26.4; 2L: -15.4), while those of 3L and 4L MoS$_2$ nanosheets are positive (3L: 4.2; 4L: 28.5). Meanwhile, the $C_{DR}$ and $C_{DG}$ values of 1L-4L MoS$_2$ nanosheets are also discrete enough for their thickness identification (Figure 4d, f and i). Therefore, 1L-15L MoS$_2$ nanosheets on 90 nm SiO$_2$/Si can be easily identified based on the measured $C_D$, $C_{DR}$, $C_{DG}$ and $C_{DB}$ values.

An interesting feature was observed in the plot of $C_D$ (or $C_{DR}$, $C_{DG}$ and $C_{DB}$) vs. layer number of MoS$_2$. For example, there is a transition of $C_D$ value between 5L and 6L nanosheets (Figure 3p and Supplementary Table S2) from negative $C_D$ at 5L (-21.8 ±0.5) to positive $C_D$ at 6L (1.0 ±0.7). In other words, compared to the 90 nm SiO$_2$/Si substrate, 1L-5L MoS$_2$ nanosheets are darker while 6L-15L MoS$_2$ nanosheets are brighter under white light illumination. In this work, the thickness of a nanosheet with a minimum positive $C_D$ value (similar for $C_{DR}$, $C_{DG}$ or $C_{DB}$) is defined as the transitional thickness ($T_C$). In this case, the $T_C$ for MoS$_2$ nanosheets in color image, grayscale image from R, G or B channel is 6L, 10L, 5L and 3L, respectively (Table 1). Therefore, the sign (positive or negative) of the $C_D$ value enables the fast determination of the thickness range of a nanosheet (i.e. below $T_C$ or above $T_C$).

Similar to MoS$_2$ nanosheets on 90 nm SiO$_2$/Si, 1L-15L MoS$_2$ nanosheets on 300 nm SiO$_2$/Si can also be reliably identified by measuring the $C_D$ value in combination with the $C_{DR}$, $C_{DG}$ and $C_{DB}$ values (Supplementary Figure S6 and Table S2).

**Optical identification of single- to quattuordecuple-layer (1L-14L) WSe$_2$ nanosheets on 90 nm SiO$_2$/Si**

The optical contrast difference can also be used to identify 1L-14L WSe$_2$ nanosheets on 90 or 300 nm SiO$_2$/Si. Figure 5a-l show the color optical images of 1L-14L WSe$_2$ nanosheets on 90 nm SiO$_2$/Si taken at the exposure time of 80 ms. The thickness of these nanosheets were confirmed by AFM measurement and low-frequency Raman spectra (Figure 5o and Supplementary Figure S7). As shown in Figure 5m, the $C_D$ values are distinguishable for 6L-14L WSe$_2$ nanosheets, but are difficult to be differentiated for 1L-5L nanosheets. As shown in Figure 5m and Supplementary Table S3, the 2L-4L WSe$_2$ nanosheets have similar $C_D$ values (2L: -59.2 ± 1.4; 3L: -62.9 ± 1.5; 4L: -55.3 ± 1.4), so do 1L and 5L WSe$_2$ nanosheets (1L: -38.4 ± 1.7; 5L: -36.8 ± 2.4). In this case, the grayscale images of R, G and B channels were used to indicate the difference among the 1L-5L WSe$_2$ nanosheets (Figure 5n). By comparing $C_{DR}$, $C_{DG}$ and $C_{DB}$ values at various layer numbers, it was found that the $C_{DR}$ values combined with $C_{DB}$ values are mostly suitable to rapidly differentiate 1L-5L WSe$_2$ nanosheets because of the sufficient gap between the $C_{DR}$ and $C_{DB}$ values of adjacent layer numbers (Supplementary Table S3). The $C_{DB}$ value of 1L WSe$_2$ nanosheet is negative (-27.6 ± 1.4) while that of 5L WSe$_2$ nanosheet is positive (16.7 ± 0.9). In addition, the $T_C$ of $C_{DB}$ values is 4L WSe$_2$ nanosheet (0.3 ± 0.8). Thus 1L, 4L and 5L WSe$_2$ nanosheets can be easily identified by reading the $C_{DB}$ values. Although 2L and 3L WSe$_2$ nanosheets have similar $C_{DG}$ values (2L: -76.8 ± 1.4; 3L: -76.5 ± 1.9), their $C_{DR}$ values (2L: -80.0 ± 0.9; 3L: -103.0 ± 0.4) are fairly discrete for thickness identification. Thus, 1L-14L WSe$_2$ nanosheets on 90 nm SiO$_2$/Si can be readily identified using the $C_D$ values in combination with $C_{DR}$ and $C_{DB}$ values. Furthermore, for 1L-14L WSe$_2$ nanosheets on 300 nm SiO$_2$/Si, their $C_D$, $C_{DR}$, $C_{DG}$ and $C_{DB}$ values are also distinguishable for fast thickness determination (Supplementary

Figure S8 and Table S3).

**Verification of layer number identification of MoS$_2$ and WSe$_2$ nanosheets on 90 nm SiO$_2$/Si**

In order to verify the accuracy of our optical method, the thicknesses of mechanically exfoliated graphene, MoS$_2$ and WSe$_2$ nanosheets on 90 nm SiO$_2$/Si were firstly identified using the measurement of $C_D$ values followed by AFM measurement to confirm it.

Taking graphene as an example, Figure 6a shows the color optical image of a graphene nanosheet. As shown in Figure 6b, the $C_D$ values measured from the red dashed rectangle shown in Figure 6a are -10.7, -57.6 and -68.3, respectively. According to the standard chart shown in Figure 2o-p and Supplementary Table S1, these $C_D$ values correspond to 1L, 5L and 6L graphene nanosheets, respectively. AFM measurement on these regions show thicknesses of 0.4, 1.7 and 2.1 nm (Figure 6c-d), respectively, consistent with the thickness of 1L, 5L and 6L graphene (Figure 2q and Table S5 in SI), respectively, confirming the accuracy of the optical identification result.

As for MoS$_2$, Figure 7a shows the color optical image of an exfoliated MoS$_2$ nanosheet, displaying three distinct color regions. $C_D$ values measured from the red dashed rectangle shown in Figure 7a are -2.5, 22.4 and 60.3, respectively, which correspond to 6L, 7L and 9L MoS$_2$ nanosheets according to the standard chart shown in Figure 3o-p and Supplementary Table S2. The corresponding thickness of the these three regions measured by AFM (Figure 7c-d) are 4.1, 4.7 and 5.9 nm, respectively, consistent with that of 6L, 7L and 9L MoS$_2$ nanosheets.

Similarly, for the optical identification of WSe$_2$ nanosheets, $C_D$ values of 9.5 and 30.8

(Figure 7f) were first obtained from two different color regions (highlighted in the red dashed rectangle shown in Figure 7e), corresponding to 7L and 8L WSe$_2$ nanosheets, respectively, by referring to the standard chart shown in Figure 5m and Supplementary Table S3. AFM measurement (Figure 7g-h) on these two regions indicates thicknesses of 4.7 and 5.4 nm, respectively, in agreement with that of 7L and 8L WSe$_2$ nanosheets (Figure 5o and Supplementary Table S5).

Besides graphene, MoS$_2$ and WSe$_2$ nanosheets, our method can also be used for the rapid and reliable identification of TaS$_2$ nanosheets on 90 nm SiO$_2$/Si. The $C_D$ values of 2L-8L, 15L to octoviguple-layer (28L) and duotriguple-layer (32L) TaS$_2$ nanosheets are discrete enough for reliable identification (Supplementary Figure S9 and Table S4). In combination with the $C_{DR}$, $C_{DG}$ and $C_{DB}$ values (Supplementary Table S4), 2L-28L and 32L TaS$_2$ nanosheets can be easily and reliably identified (Supplementary Figures S9-S10), indicating the generalizability of our method in the thickness identification of 2D nanosheets.

The measurement of $C_D$ is affected by the intensity of illumination, thickness of SiO$_2$ film, and exposure time. In our work, the thickness of SiO$_2$ film and exposure time are fixed. Therefore, the measurement error (characterized by the standard deviation, SD) likely arises from the fluctuation of illumination intensity, which is manually adjusted in our optical microscope (Supplementary Figure S11). Nevertheless, the measurement error is much smaller than the difference between $C_D$ values of adjacent layers. In other words, the difference among $C_D$ values is sufficient for thickness determination. In addition, the transition thickness ($T_C$) of optical contrast difference is related to the type of material, the thickness of SiO$_2$ film, as well as the color of optical image. As shown in Table 1, the $T_C$ values of $C_D$, $C_{DR}$, $C_{DG}$ and $C_{DB}$ follow the order of $C_{DR} > C_D \geq C_{DG} > C_{DB}$ for 1L-15L MoS$_2$ and 1L-14L WSe$_2$ nanosheets on 90 and 300 nm SiO$_2$/Si. In the case of TaS$_2$ nanosheets on 90 nm SiO$_2$/Si, the $T_C$ values of $C_D$, $C_{DR}$ and $C_{DG}$ follow the order of $C_D > C_{DG} > C_{DR}$ (Table 1). In terms of graphene nanosheets on 90 and 300 nm SiO$_2$/Si, the $T_C$ values of $C_D$, $C_{DR}$, $C_{DG}$ and $C_{DB}$ are much larger compared to those of MoS$_2$, WSe$_2$ and TaS$_2$ nanosheets (Table 1 and Supplementary Figures S12-13). The variation of $T_C$ for different materials might be attributed to their intrinsic properties, such as refractive index.

## CONCLUSION

In summary, a universal optical method has been developed for simple, rapid and reliable identification of 1L-15L 2D nanosheets, including graphene, MoS$_2$, WSe$_2$ and TaS$_2$, on 90 and 300 nm SiO$_2$/Si. By processing the color optical images and the grayscale images of R, G and B channels, the optical contrast differences between 2D nanosheets and SiO$_2$/Si were measured using ImageJ and plotted as standard charts to guide the layer number identification. The transition of $C_D$, $C_{DR}$, $C_{DG}$, and $C_{DB}$ values can be used as clear mark for quick identification of layer number. Neither complex calculation nor special instrument is required in our method, making it applicable for any labs equipped with standard optical microscope and digital camera. Our simple optical identification method will facilitate the fundamental study and practical applications of 2D nanomaterials, and accelerate their progress towards future commercialization. Furthermore, our method potentially expands the capability of optical microscope in study of nanomaterials and applications of nanotechnology.

**METHODS AND MATERIALS**

*Mechanical exfoliation of 2D nanosheets (graphene, MoS$_2$, WSe$_2$ and TaS$_2$ nanosheets).* Natural graphite (NGS Naturgraphit GmbH, Germany), MoS$_2$ crystals (SPI Supplies, USA), WSe$_2$ and TaS$_2$ crystals (Nanoscience Instruments, Inc., USA) were used for preparation of mechanically-exfoliated 2D nanosheets, respectively, which then were deposited onto the freshly cleaned 90 and 300 nm SiO$_2$-coated Si substrates (90 and 300 nm SiO$_2$/Si).

*Capture of optical images of 2D nanosheets.* The bright-field optical microscope (Eclipse LV100D with a 100×, 0.9 numerical aperture (NA) objective, Nikon) was used to locate and image the 2D nanosheets. A lamphouse (LV-LH50PC) equipped with high-intensity halogen lamp (12V-50W) was used as light source. The intensity of light source was adjusted by turning the brightness control knob to level 9 (Supplementary Figure S11). For graphene and MoS$_2$ nanosheets on 90 nm SiO$_2$/Si, the optical images were captured at the exposure time of 20, 40, 60, 80, 100, 120, 140, 160, 180, 200, 250 and 300 ms, respectively. For WSe$_2$ and TaS$_2$ nanosheets on 90 nm SiO$_2$/Si, the optical images were captured at the exposure time of 80 ms. For various 2D nanosheets on 300 nm SiO$_2$/Si, the optical images were captured at the exposure time of 50 ms. A DS camera head (DS-Fi1) with a digital camera control unit (DS-U3) was used to capture color optical images of 2D nanosheets at the resolution of 1280 × 960 pixels. The imaging software is NIS-Elements F (version 4.00.06) and the white balance is calibrated as R/B=1.23:1.24 (Supplementary Figure S11). In order to give quantitative and statistic characterization of the layer numbers of 2D nanosheets, a large amount of graphene,

MoS$_2$, WSe$_2$ and TaS$_2$ flakes with layer numbers ranged from 1L to 32L, prepared by the mechanical cleavage technique, was imaged by optical microscopy and analyzed by using our optical method. For 1L to 10L 2D nanosheets, at least 5 samples were collected for measurement. For 2D nanosheets thicker than 11L, usually 3 samples were collected for measurement.

***Optical contrast difference measurement of color optical images and grayscale images of R, G and B channels by using ImageJ.*** The color optical images of 2D nanosheets were processed by the ImageJ (version 1.46p, National Institutes of Health, USA). For the color image (RGB image), the contrast value of each pixel ($C_V$), i.e. brightness value, is calculated using the following equation,

$$C_V = (C_{VR} + C_{VG} + C_{VB})/3 \qquad (5)$$

where, $C_{VR}$, $C_{VG}$ and $C_{VB}$ are the R, G and B values per pixel in color image, respectively. 0 means darkest and 255 means brightest.

The grayscale images of R, G and B channels were extracted by using the "Split Channels" command from "Image > Color > Split Channels" in the menu bar, where 0 means darkest and 255 means brightest. In the grayscale image, we can drag the left button of mouse to draw a rectangular box across the 2D nanosheet and then press "K" to obtain the contrast profile of the selected area. In the plot of contrast profile, click "List" to show the contrast values of the 2D nanosheet and SiO$_2$ substrate.

The detailed optical contrast difference ($C_D$, $C_{DR}$, $C_{DG}$ and $C_{DB}$) values of 2D nanosheets are listed in Supplementary Table S1-S4.

***Thickness measurement of 2D nanosheets by AFM.*** AFM (Dimension ICON with NanoScope V controller, Bruker, USA) was used to confirm the layer number of 2D nanosheets by measuring the film thickness in tapping mode in air. The thickness values of 2D nanosheets are listed in Supplementary Table S5.

***Raman measurement of 2D nanosheets.*** Analysis of the $MoS_2$ and $WSe_2$ nanosheets by low-frequency Raman spectroscopy was carried out at room temperature using a micro-Raman spectrometer (Horiba-JY T64000) equipped with a liquid nitrogen cooled charge-coupled device. The measurements were conducted in a backscattering configuration excited with a solid state green laser ($\lambda = 532$ nm). A reflecting Bragg grating (OptiCrate) followed by another ruled reflecting grating was used to filter the laser side bands, as such ~8 $cm^{-1}$ limit of detection was achieved using most solid state or gas laser lines. All spectra were collected through a 100× objective and dispersed by a 1800 g/mm grating under a triple subtractive mode with a spectra resolution of 1 $cm^{-1}$. The Raman spectra were calibrated by using the peak of Si substrate (520 $cm^{-1}$). The laser power at the sample surface was less than 1.5 mW for $MoS_2$ and 0.3 mW for $WSe_2$, respectively.

Analysis of graphene nanosheets by Raman spectroscopy was carried out on a WITec CRM200 confocal Raman microscopy system with the excitation line of 488 nm and an air cooling charge coupled device (CCD) as the detector (WITec Instruments Corp, Germany).

**Acknowledgement.**

This work was supported by AcRF Tier 1 (RG 61/12) and Start-Up Grant (M4080865.070.706022) in Singapore. This research is also funded by the Singapore National Research Foundation and the publication is supported under the Campus for Research Excellence And Technological Enterprise (CREATE) programme.

**Corresponding Author**

*E-mail: hzhang@ntu.edu.sg, hzhang166@yahoo.com.

Phone: +65-6790-5175. Fax: +65-6790-9081

Website: http://www.ntu.edu.sg/home/hzhang/

*Conflict of Interest:* The authors declare no competing financial interest.

*Supporting Information Available.* Raman spectra of 1L-15L graphene, $MoS_2$ and $WSe_2$ nanosheets. Optical images and contrast difference plots of 1L-15L graphene, $MoS_2$ and $WSe_2$ nanosheets on 300 nm $SiO_2$/Si. Optical images and contrast difference plots of 2L-28L and 32L $TaS_2$ nanosheets on 90 nm $SiO_2$/Si. The detailed optical contrast difference ($C_D$, $C_{DR}$, $C_{DG}$ and $C_{DB}$) and height values of 2D nanosheets. These materials are available free of charge *via* the Internet at http://pubs.acs.org.

**Figure Caption**

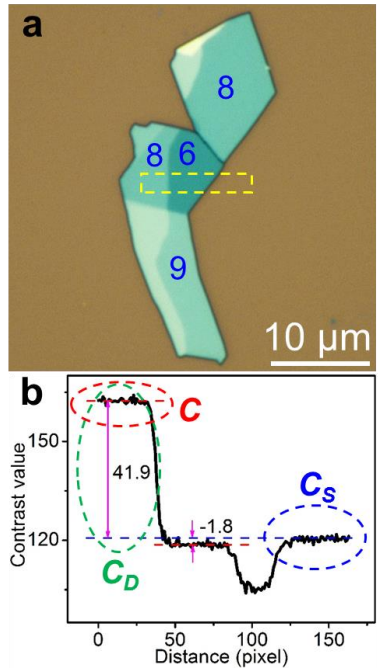

**Figure 1.** (a) Color optical image of a $MoS_2$ flake deposited on 90 nm $SiO_2$/Si. The digitals shown in (a) indicate the layer numbers of $MoS_2$ nanosheets. (b) Contrast profile of the dashed rectangle shown in (a). $C_S$: contrast of 90 nm $SiO_2$/Si. $C$: contrast of $MoS_2$ nanosheet. $C_D$: the contrast difference between $MoS_2$ nanosheet and 90 nm $SiO_2$/Si.

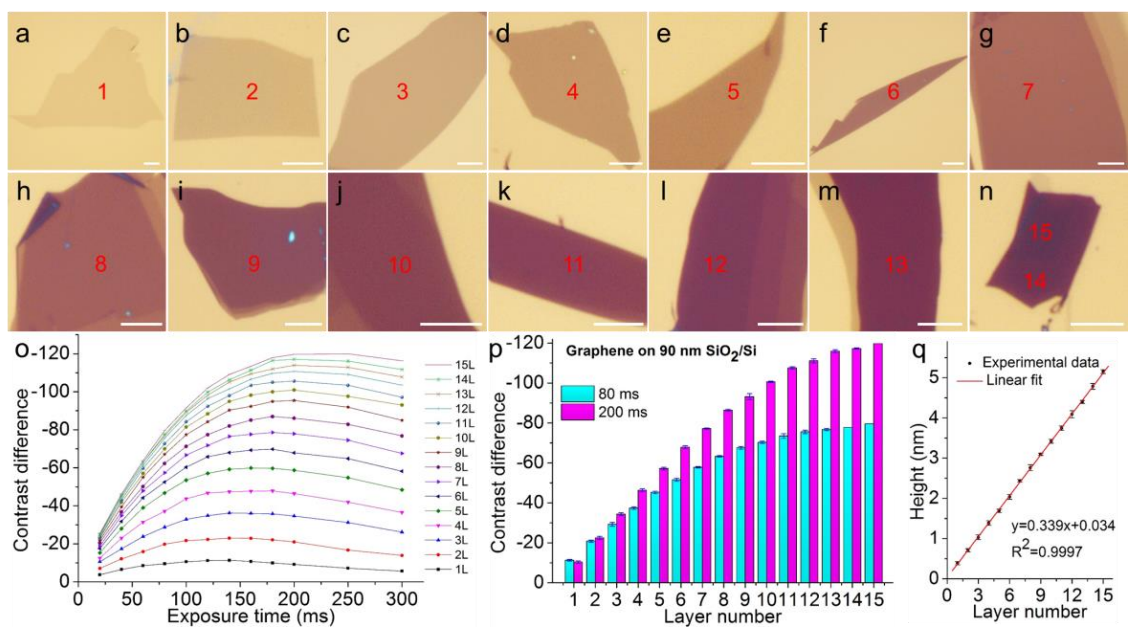

**Figure 2**. (a-n) Color optical images of 1L-15L graphene nanosheets on 90 nm SiO$_2$/Si. The scale bars shown in (a-n) are 10 μm. The digitals shown in (a-n) indicate the layer numbers of corresponding graphene nanosheets. (o) Plot of measured $C_D$ values of 1L-15L graphene nanosheets on 90 nm SiO$_2$/Si at the exposure time of 20, 40, 60, 80, 100, 120, 140, 160, 180, 200, 250 and 300 ms, respectively. (p) Plot of measured $C_D$ values of 1L-15L graphene nanosheets on 90 nm SiO$_2$/Si at the exposure time of 80 and 200 ms, respectively. (q) The thickness of 1L-15L graphene nanosheets measured by AFM.

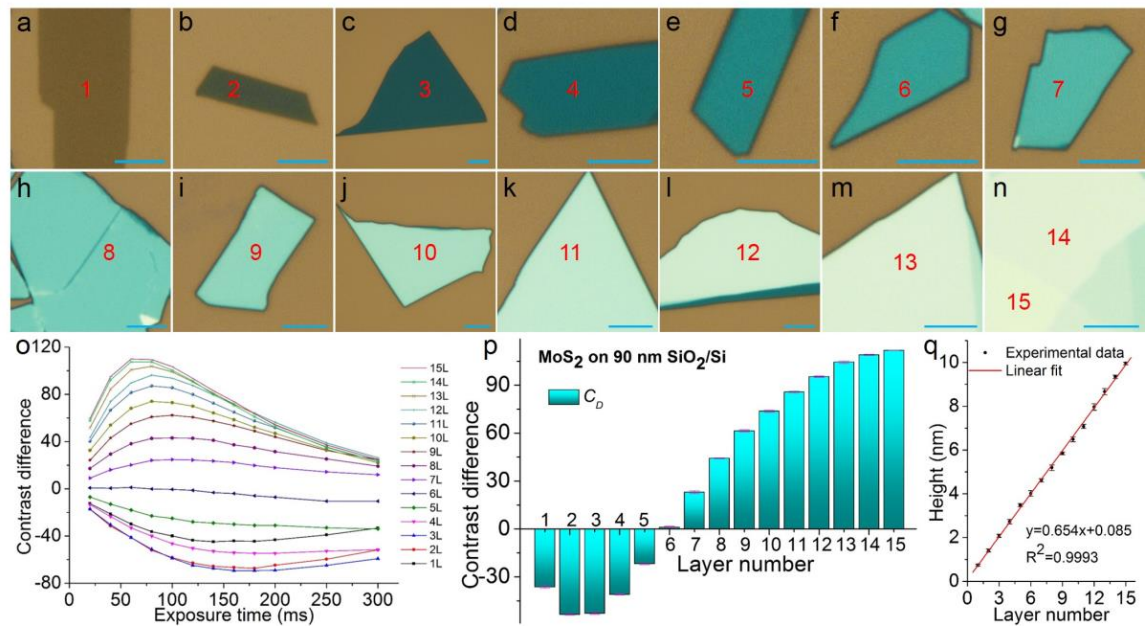

**Figure 3**. (a-n) Color optical images of 1L-15L MoS$_2$ nanosheets on 90 nm SiO$_2$/Si. The scale bar is 5 μm for each image. The digitals shown in (a-n) indicate the layer numbers of corresponding MoS$_2$ nanosheets. (o) Plot of measured $C_D$ values of 1L-15L MoS$_2$ nanosheets on 90 nm SiO$_2$/Si at exposure time of 20, 40, 60, 80, 100, 120, 140, 160, 180, 200, 250 and 300 ms, respectively. (p) Plot of $C_D$ values of 1L-15L MoS$_2$ on 90 nm SiO$_2$/Si at exposure time of 80 ms. (q) The thickness of 1L-15L MoS$_2$ nanosheets measured by AFM.

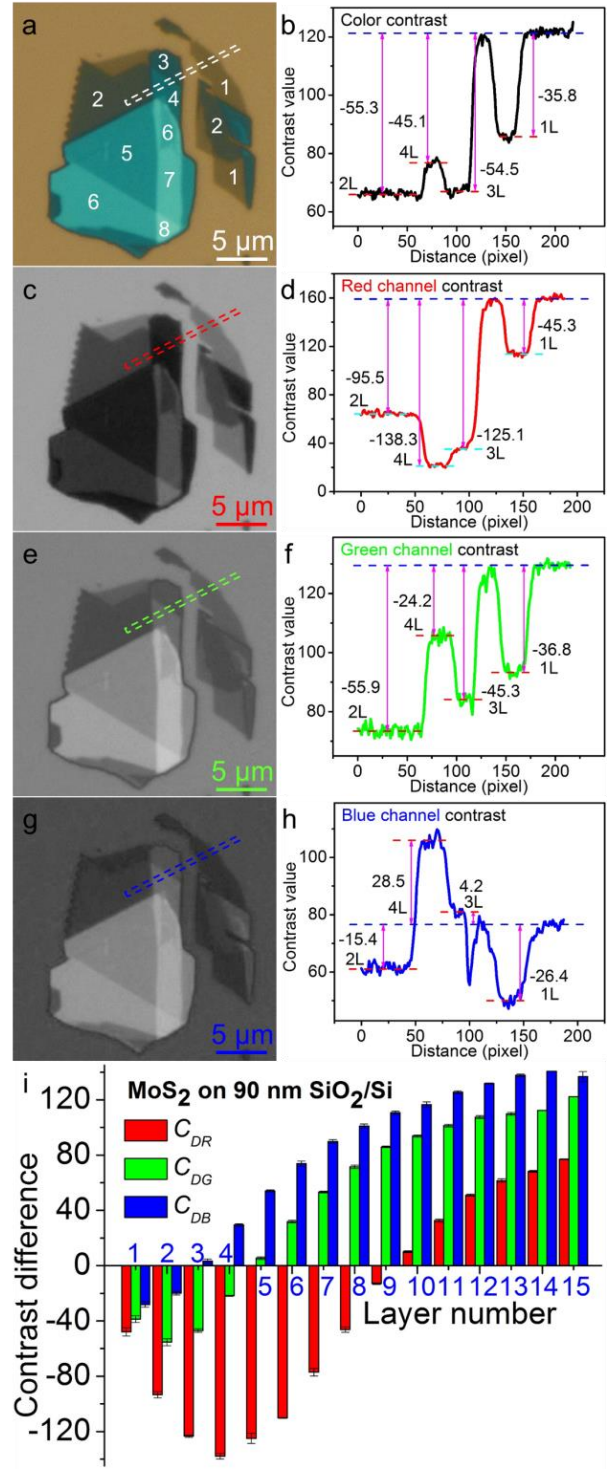

**Figure 4**. Color optical (a) and grayscale images of the R (c), G (e) and B (g) channels of MoS$_2$ flake on 90 nm SiO$_2$/Si. The digitals shown in (a) indicate the layer numbers of corresponding MoS$_2$ nanosheets. The corresponding contrast profiles of color optical (b)

and grayscale images of the R (d), G (f) and B (h) channels of MoS$_2$ flake are obtained from the dashed rectangles shown in (a), (c), (e) and (g), respectively. (i) Plot of $C_{DR}$, $C_{DG}$ and $C_{DB}$ values of 1L-15L MoS$_2$ nanosheets on 90 nm SiO$_2$/Si.

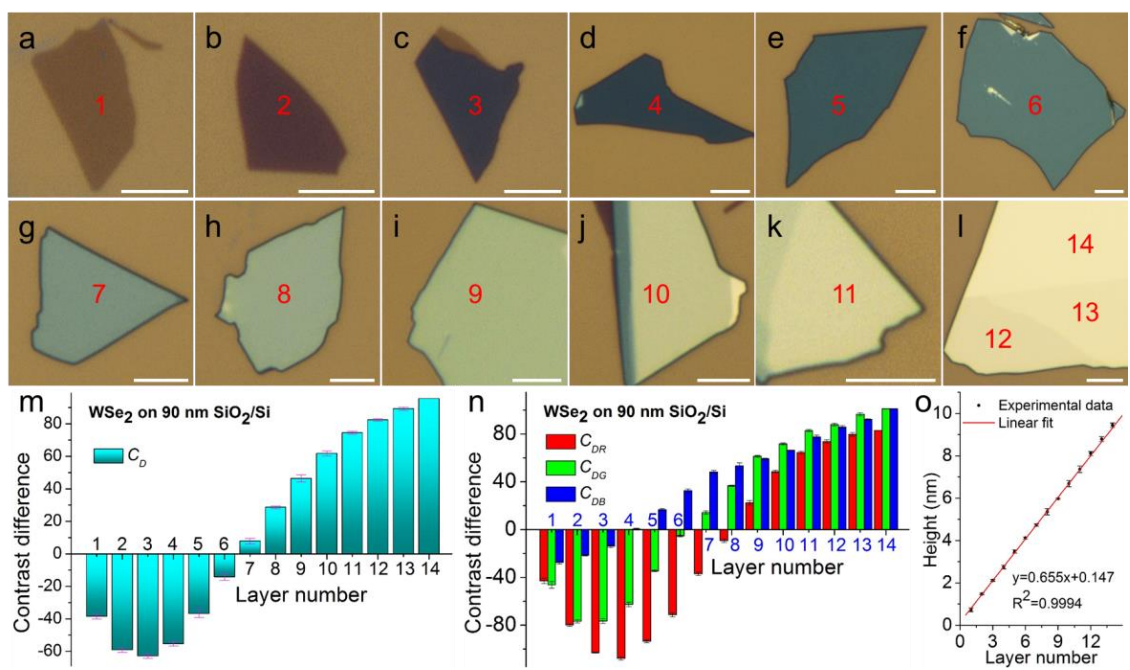

**Figure 5**. (a-l) Color optical images of 1L-14L WSe$_2$ nanosheets on 90 nm SiO$_2$/Si. The scale bars shown in (a-i) are 5 μm. The digitals shown in (a-l) indicate the layer numbers of corresponding WSe$_2$ nanosheets. (m-n) Plots of (m) $C_D$ and (n) $C_{DR}$, $C_{DG}$ and $C_{DB}$ values of 1L-14L WSe$_2$ nanosheets on 90 nm SiO$_2$/Si at the exposure time of 80 ms. (o) The thickness of 1L-14L WSe$_2$ nanosheets measured by AFM.

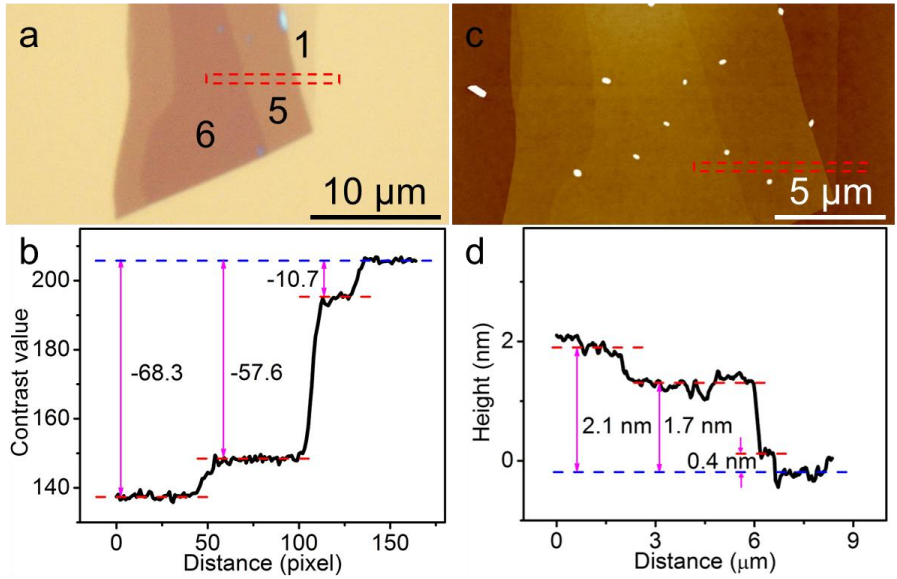

**Figure 6**. Color optical (a) and AFM height (c) images of graphene flake on 90 nm SiO$_2$/Si. The corresponding contrast difference (b) and height (d) profiles are obtained from the dashed rectangles shown in (a) and (c), respectively. The digitals shown in (a) indicate the layer numbers of corresponding graphene nanosheets.

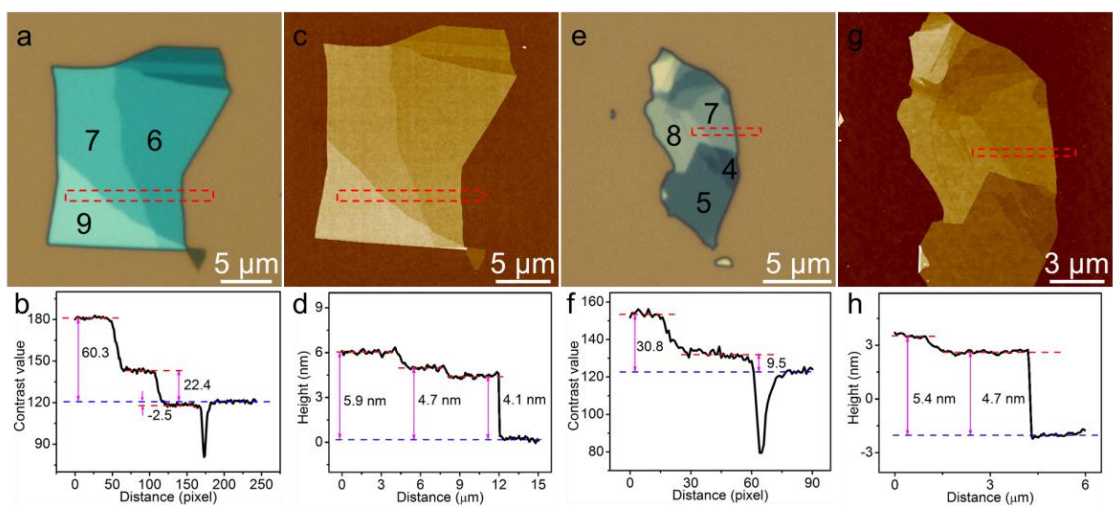

**Figure 7**. Color optical images of MoS$_2$ (a) and WSe$_2$ (e) nanosheets on 90 nm SiO$_2$/Si and the corresponding contrast profiles (b and f) obtained from the dashed rectangles in (a) and (e), respectively. AFM height images of MoS$_2$ (c) and WSe$_2$ (g) nanosheets on 90 nm SiO$_2$/Si, and the corresponding height profiles (d and h) obtained from the dashed rectangles in (c) and (g), respectively. The digitals shown in (a) and (e) indicate the layer numbers of corresponding MoS$_2$ and WSe$_2$ nanosheets, respectively.

**Table 1**. **The transition thickness of 2D nanosheets with minimum positive optical contrast difference on 90 and 300 nm SiO$_2$/Si.**

|  | 90 nm SiO$_2$/Si | | | | 300 nm SiO$_2$/Si | | | |
| --- | --- | --- | --- | --- | --- | --- | --- | --- |
|  | $C_D$ | $C_{DR}$ | $C_{DG}$ | $C_{DB}$ | $C_D$ | $C_{DR}$ | $C_{DG}$ | $C_{DB}$ |
| Graphene | 43L < Tc < 52L* | 54L < Tc* | 43L < Tc < 52L* | 30L < Tc < 37L* | 46L < Tc < 51L* |  | ~ 46L* |  |
| MoS$_2$ | 6L | 10L | 5L | 3L | 8L | 16L | 4L | 1L |
| WSe$_2$ | 7L | 9L | 7L | 4L | 8L | 14L | 5L | 1L |
| TaS$_2$ | 27L | 22L | 25L | 3L |  |  |  |  |

*see Figures S12-S13 in Supplementary Information for the detailed information.

# Supporting Information

# Rapid and reliable thickness identification of two-dimensional nanosheets using optical microscopy

*Hai Li,[1,†] Jumiati Wu,[1,†] Xiao Huang,[1] Gang Lu,[1] Jian Yang,[1] Xin Lu,[2] Qihua Xiong,[2] Hua Zhang[1,*]*

[1]School of Materials Science and Engineering, Nanyang Technological University, 50 Nanyang Avenue, Singapore 639798, Singapore

[2]Division of Physics and Applied Physics, School of Physical and Mathematical Sciences, Nanyang Technological University, 21 Nanyang Link, Singapore 637371, Singapore

*To whom correspondence should be addressed.

Phone: +65-6790-5175. Fax: +65-6790-9081

E-mail: hzhang@ntu.edu.sg

Website: http://www.ntu.edu.sg/home/hzhang/

[†]These authors contributed equally to this work.

## 1. Raman spectra of graphene nanosheets on 90 and 300 nm SiO₂/Si.

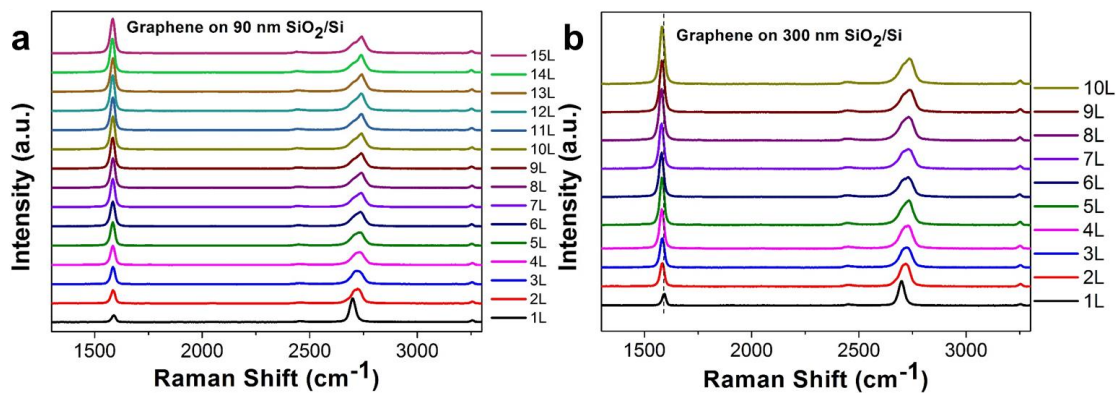

**Figure S1**. Raman spectra of (a) single- to quindecuple-layer (1L-15L) graphene nanosheets on 90 nm SiO₂/Si, and (b) 1L to decuple-layer (10L) graphene nanosheets on 300 nm SiO₂/Si in the range of 1300-3300 cm$^{-1}$.

## 2. Optical identification of 1L-15L graphene nanosheets on 90 nm SiO₂/Si by using $C_{DR}$, $C_{DG}$ and $C_{DB}$ values.

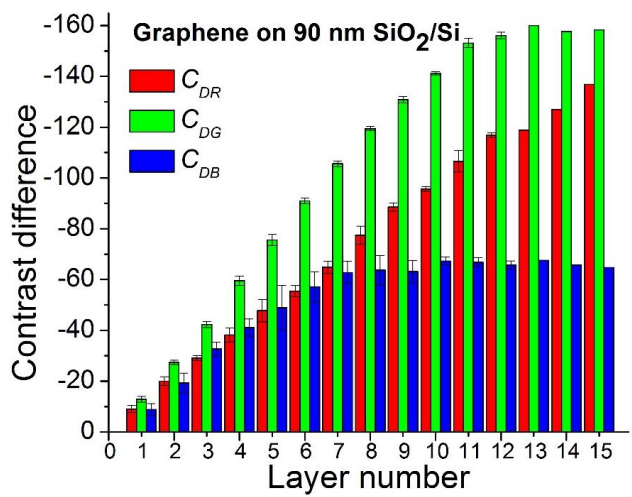

**Figure S2**. Plot of $C_{DR}$, $C_{DG}$ and $C_{DB}$ values vs. layer number of 1L-15L graphene nanosheets on 90 nm SiO₂/Si. The original color optical images were taken at exposure

time of 200 ms.

In Figure S2, it can be seen that the $C_{DG}$ values of 1L to undecuple-layer (11L) graphene nanosheets decrease almost linearly with thickness and thus can be used for the thickness identification of graphene. However, for the thicker graphene nanosheets, i.e. tredecuple-layer (13L) to 15L, the $C_{DR}$ values are more suitable for the graphene thickness identification. In addition, the $C_{DB}$ values are distinguishable for 1L to triple-layer (3L) graphene, but show less difference from quadruple-layer (4L) to 15L nanosheets (considering the error bars) and thus are not suitable for thickness identification of 4L-15L graphene.

**3. Optical identification of single- to tredecuple-layer (1L-13L) graphene nanosheets on 300 nm SiO$_2$/Si**

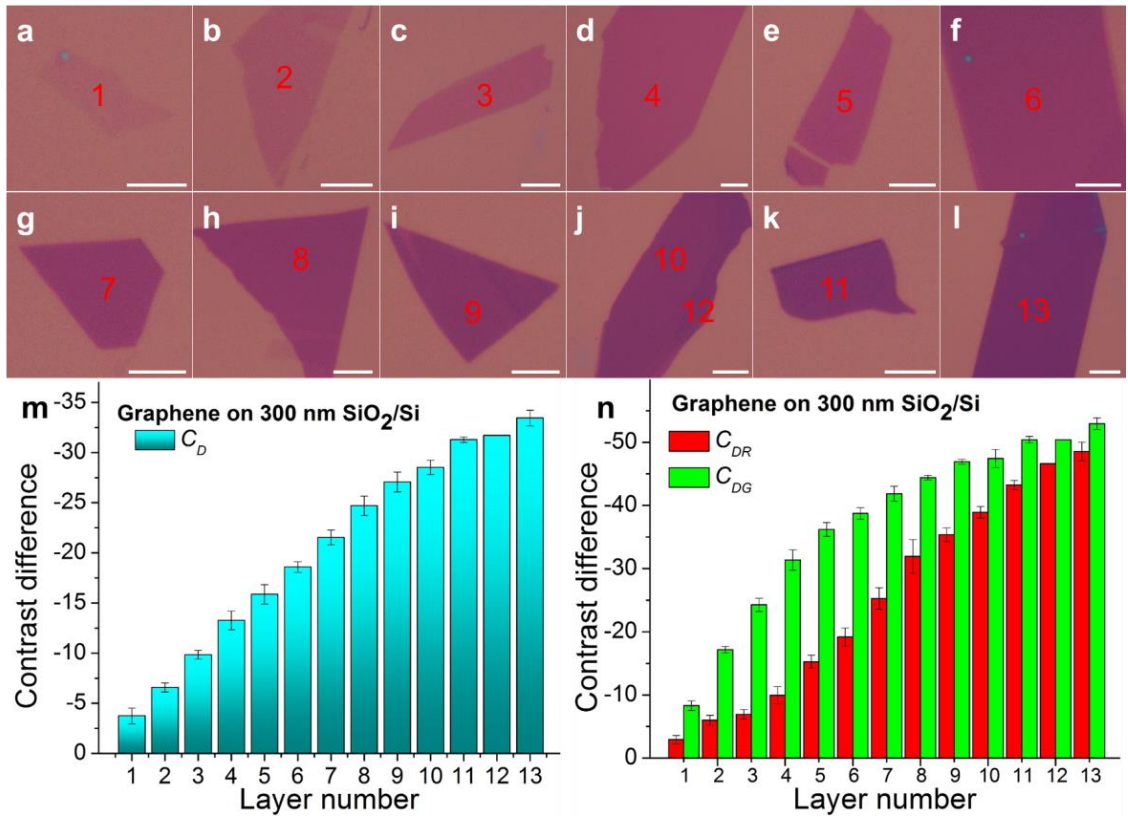

**Figure S3**. (a-l) Color optical images of 1L-13L graphene nanosheets on 300 nm $SiO_2$/Si taken at the exposure time of 50 ms. The scale bars shown in (a-l) are 5 μm. The digitals shown in (a-l) indicate the layer numbers of the corresponding graphene nanosheets. (m-n) Plots of (m) $C_D$ values and (n) $C_{DR}$ and $C_{DG}$ values of 1L-13L graphene nanosheets on 300 nm $SiO_2$/Si.

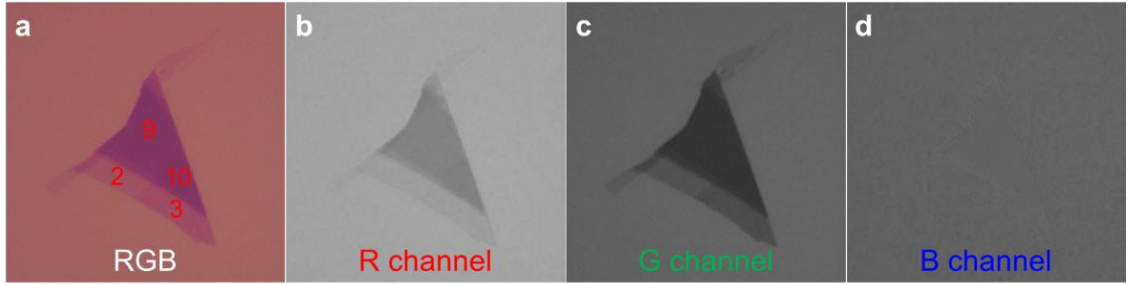

**Figure S4**. (a) Color optical and (b-d) grayscale images of R, G and B channels of a graphene flake on 300 nm $SiO_2$/Si. It is difficult to distinguish the graphene flake and substrate in the grayscale image of B channel shown in (d). The digitals shown in (a) indicate the layer numbers of the corresponding graphene nanosheets.

By using 300 nm $SiO_2$/Si, the $C_D$ values can be used to rapidly and reliably identify the graphene nanosheets from 1L to octuple-layer (8L) (Figure S3m), while the $C_{DR}$ values are suitable for identification of 5L-13L graphene nanosheets (Figure S3n). The $C_{DG}$ values are also suitable for identification of 1L-5L graphene nanosheets (Figure S3n). However, the grayscale image of the B channel of a graphene nanosheet on 300 nm $SiO_2$/Si does not show obvious contrast between graphene and the substrate, thus the $C_{DB}$ values cannot be used for the identification of graphene thickness (Figure S4d). Therefore, 1L-13L graphene nanosheets on 300 nm $SiO_2$/Si can be readily identified using the $C_D$ values in combination with the $C_{DR}$ and $C_{DG}$ values.

## 4. Low-frequency and normal Raman spectra of 1L-15L MoS$_2$ nanosheets.

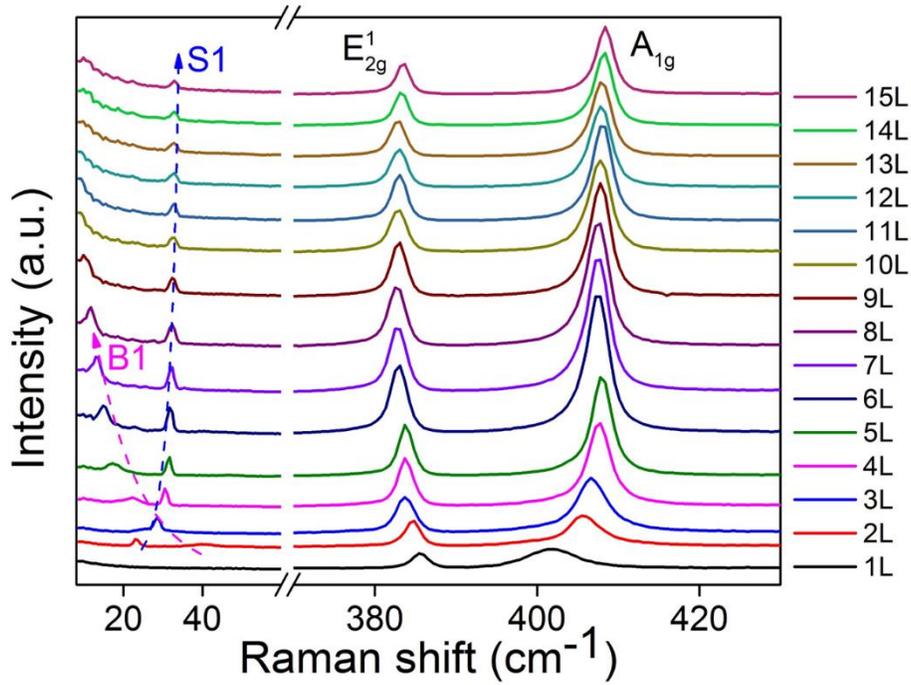

**Figure S5**. Raman spectra of 1L-15L MoS$_2$ nanosheets in the range of 8-430 cm$^{-1}$. Dashed curves are guides for the eyes. S1 and B1 represent the shear and layer breathing (LB) modes, respectively. A$_{1g}$ peaks can be used to identify 1L-4L MoS$_2$ nanosheets. However, the dependency of A$_{1g}$ peak position on the layer number becomes less obvious for 5L and thicker nanosheets. In addition, the LB mode (B1) peaks can be used to differentiate 1L to nonuple-layer (9L) MoS$_2$ nanosheets. Note that B1 peaks of 10L-15L nanosheets are out of the detection range of our instrument.

## 5. Optical identification of 1L-15L MoS$_2$ nanosheets on 300 nm SiO$_2$/Si.

1L-15L MoS$_2$ nanosheets on 300 nm SiO$_2$/Si can also be identified by measuring the $C_D$ value in combination with the $C_{DR}$, $C_{DG}$ and $C_{DB}$ values (Figure S6). The $C_D$ values can be used to rapidly and reliably identify septuple-layer (7L) to 15L MoS$_2$ nanosheets, but are less distinguishable among 1L-6L nanosheets (Figure S6n). In contrast to MoS$_2$ nanosheets on 90 nm SiO$_2$/Si (Figure 2 in the main text), the transition of $C_D$ occurs between 7L and 8L MoS$_2$ nanosheets on 300 nm SiO$_2$/Si, where 7L MoS$_2$ gives a negative $C_D$ (-4.4 ± 0.2) and 8L MoS$_2$ gives a positive $C_D$ (4.7 ± 0.3). The transition of $C_D$ can be used as a mark to quickly determine an MoS$_2$ nanosheet thicker or thinner than 8L on 300 nm SiO$_2$/Si. Meanwhile, the $C_{DR}$, $C_{DG}$ and $C_{DB}$ values are used to identify 1L-6L MoS$_2$ nanosheets, which are difficult to be distinguished by the measurement of $C_D$. For example, 1L-3L MoS$_2$ nanosheets can be differentiated by reading the $C_{DR}$ values (Figure S6o and Table S2), whereas 4L-6L MoS$_2$ can be distinguished based on $C_{DG}$ values (Figure S6o and Table S2). The $T_C$ of $C_{DG}$ is 4L MoS$_2$ nanosheet (6.0 ± 1.3), that is, the $C_{DG}$ values of 1L-3L MoS$_2$ nanosheets are negative while those of 4L-6L MoS$_2$ nanosheets are positive (See Table S2 for detailed information). The highest absolute value of $C_{DR}$ was found at 6L MoS$_2$ nanosheet. After that, the absolute $C_{DR}$ values of 7L-15L MoS$_2$ nanosheets decrease linearly with the thickness. Therefore, 1L-15L MoS$_2$ nanosheets on 300 nm SiO$_2$/Si can be readily identified using the $C_D$ values in combination with the $C_{DR}$ and $C_{DG}$ values.

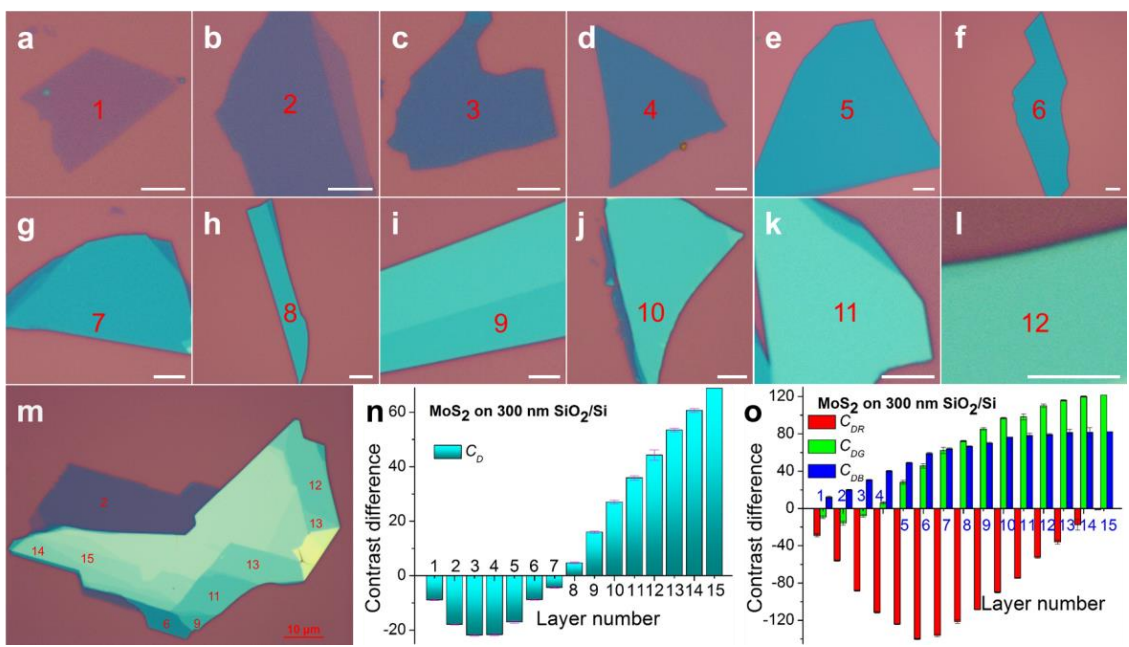

**Figure S6.** (a-m) Color optical images of 1L-15L MoS$_2$ on 300 nm SiO$_2$/Si taken at 50 ms. The scale bars are 5 μm for images a-l and 10 μm for image m, respectively. The digitals shown in (a-m) indicate the layer numbers of the corresponding MoS$_2$ nanosheets. (n-o) Plots of (n) $C_D$ values and (o) $C_{DR}$, $C_{DG}$ and $C_{DB}$ values of 1L-15L MoS$_2$ nanosheets on 300 nm SiO$_2$/Si.

## 6. Low-frequency Raman spectra of 1L to quattuordecuple-layer (14L) WSe$_2$ nanosheets.

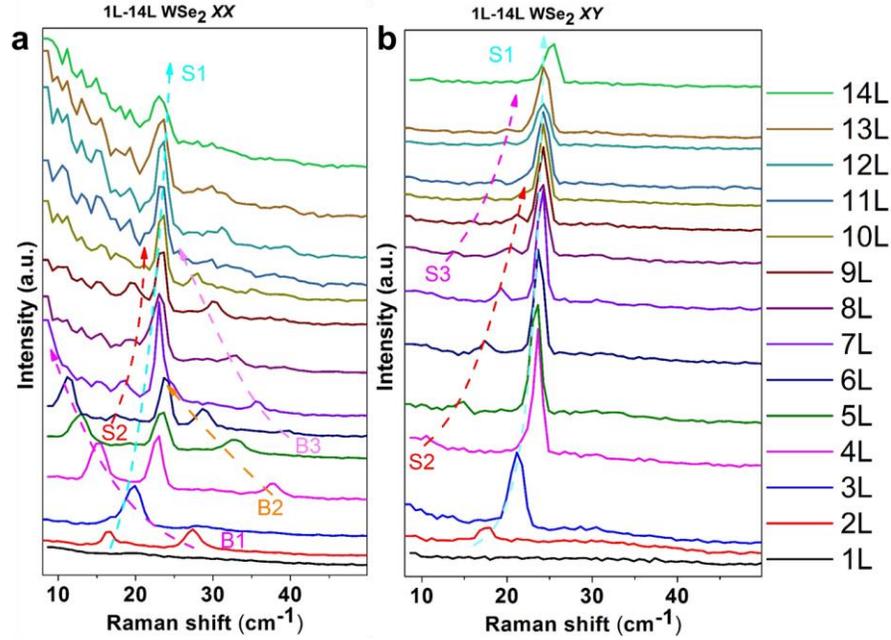

**Figure S7**. Raman spectra of 1L−14L WSe$_2$ nanosheets in the range of 8-50 cm$^{-1}$ measured under *XX* (a) and *XY* (b) polarizations, respectively. Dashed curves in (a-b) are guides for the eyes. S1-S3 and B1-B3 represent the shear and layer breathing (LB) modes, respectively. The peaks of shear and LB modes can be used to identify 1L-14L WSe$_2$ nanosheets. It is clear that B1, B2 and B3 peaks are red-shifted for 2L-6L, 4L-7L and 6L-11L, respectively. S1 peak is blue-shifted from 2L to 4L WSe$_2$ but its position becomes less affected by layer number for nanosheets thicker than 4L. On the other hand, the blue-shift of S2 and S3 peaks has been observed for 4L-10L and 8L-13L WSe$_2$, respectively. The low-frequency Raman spectra of 1L-14L WSe$_2$ nanosheets have been used to confirm that the optical identification result (Figure 5 in the main text) is correct.

## 7. Optical identification of 1L-14L WSe$_2$ nanosheets on 300 nm SiO$_2$/Si.

For the 1L-14L WSe$_2$ nanosheets on 300 nm SiO$_2$/Si (Figure S7a-m), the $T_C$ of $C_D$ is 8L (Table 1 and Table S3) and the $C_D$ values of 8L-14L WSe$_2$ can be used to rapidly and reliably distinguish them (Figure S8n). As for 1L-7L WSe$_2$ nanosheets, the $C_{DR}$, $C_{DG}$ and $C_{DB}$ values are used to determine their thicknesses (Table S3 and Figure S8o). Therefore, 1L-14L WSe$_2$ nanosheets on 300 nm SiO$_2$/Si can be readily identified using the $C_D$ values in combination with the $C_{DR}$ and $C_{DG}$ values.

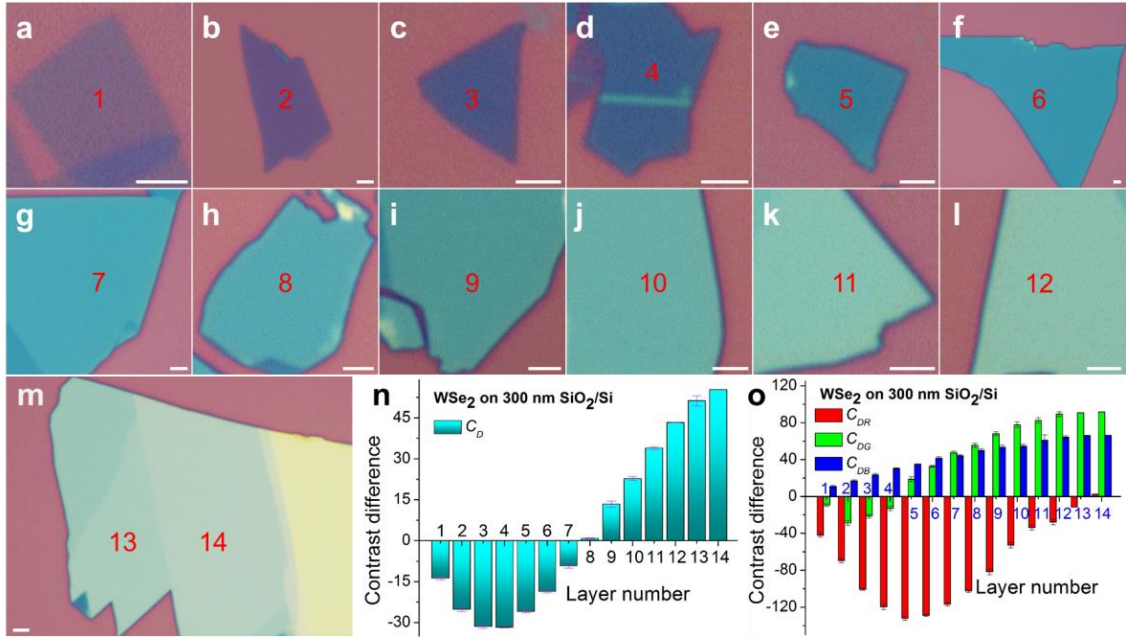

**Figure S8.** (a-m) Color optical images of 1L-14L WSe$_2$ on 300 nm SiO$_2$/Si taken at 50 ms. The scale bars shown in (a-m) are 2 μm. The digitals shown in (a-m) indicate the layer numbers of the corresponding WSe$_2$ nanosheets. (n-o) Plots of (n) $C_D$ values and (o) $C_{DR}$, $C_{DG}$ and $C_{DB}$ values of 1L-14L WSe$_2$ nanosheets on 300 nm SiO$_2$/Si.

## 8. Optical identification of 2L to octoviguple-layer (28L) and duotriguple-layer (32L) TaS$_2$ nanosheets on 90 nm SiO$_2$/Si.

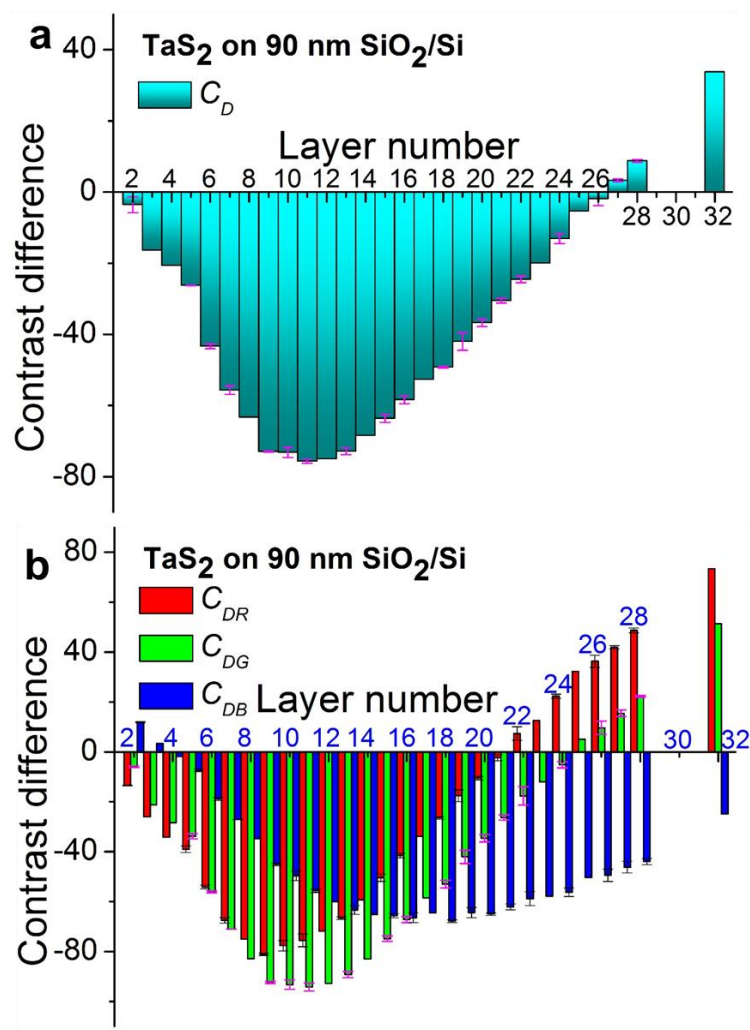

**Figure S9**. Plots of (a) $C_D$ values and (b) $C_{DR}$, $C_{DG}$ and $C_{DB}$ values of 2L-28L and 32L TaS$_2$ nanosheets on 90 nm SiO$_2$/Si. Original color images were taken at the exposure time of 80 ms.

As shown in Figure S9a, the $C_D$ values of 2L-8L, 15L-28L and 32L TaS$_2$ nanosheets are discrete enough for reliable identification (Table S4). The 9L-14L TaS$_2$ nanosheets can be identified by the $C_{DR}$, $C_{DG}$ and $C_{DB}$ values (Figure S9b). The $Tc$ of $C_{DR}$ and $C_{DG}$ is 22L and 25L (Table 1 and Table S4), respectively. Interestingly, the $C_{DB}$ values of 2L-3L TaS$_2$ nanosheets are positive, and show two $Tc$ values with one at 3L and another one probably larger than 32L. Therefore, the combination of $C_D$, $C_{DR}$, $C_{DG}$ and $C_{DB}$ values enables the easy and reliable identification of 2L-28L and 32L TaS$_2$ nanosheets on 90 nm SiO$_2$/Si.

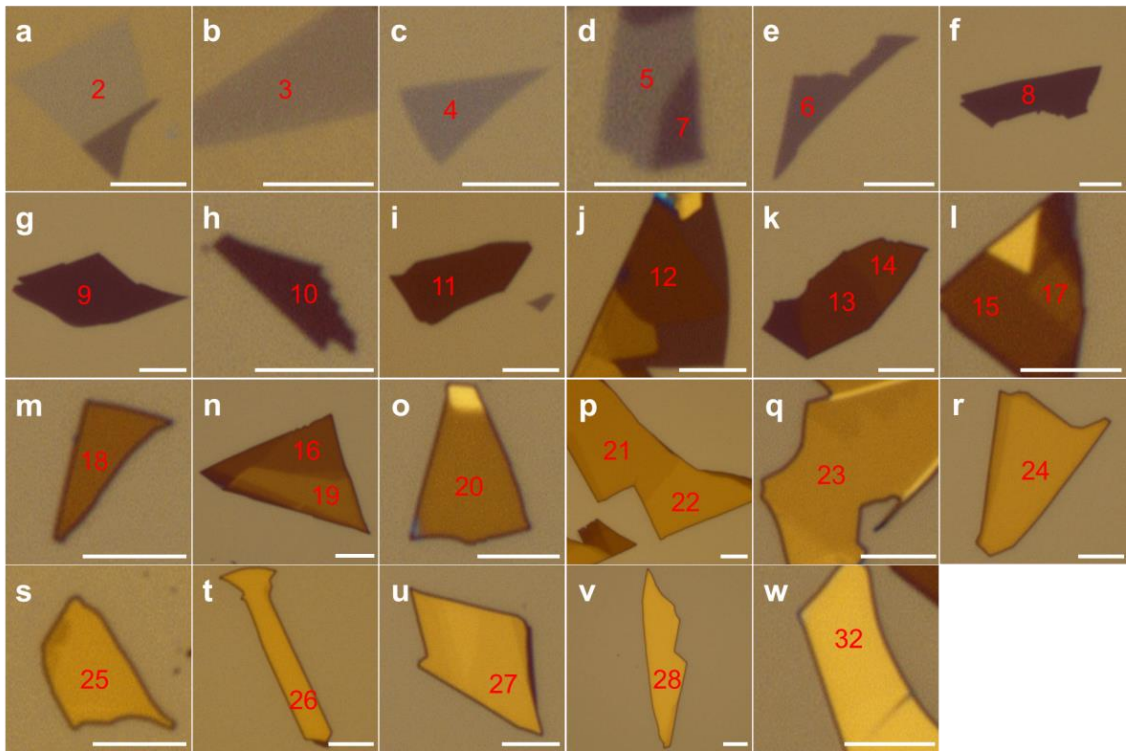

**Figure S10**. (a-w) Color optical images of 1L-28L and 32L TaS$_2$ nanosheets on 90 nm SiO$_2$/Si. The scale bars are 5 μm. The digitals shown in (a-w) indicate the layer numbers of the corresponding TaS$_2$ nanosheets.

**9. Adjustment of light intensity and software configuration of our optical microscope.**

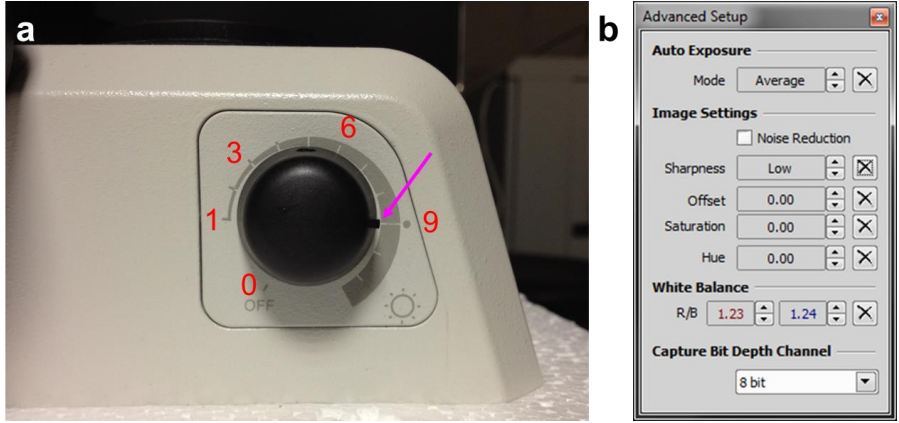

**Figure S11**. (a) The intensity of light was adjusted by turning the brightness control knob to level 9. (b) Configuration of software for capturing color optical images in the present study.
S40

## 10. Optical contrast difference ($C_D$, $C_{DR}$, $C_{DG}$ and $C_{DB}$) values of 2D nanosheets.

**Table S1.** The optical contrast difference ($C_D$, $C_{DR}$, $C_{DG}$ and $C_{DB}$) values of graphene nanosheets with different layer numbers on 90 and 300 nm SiO$_2$/Si. Optical images of graphene nanosheets on 90 and 300 nm SiO$_2$/Si were taken at the exposure time of 200 and 50 ms, respectively.

|  | 90 nm SiO$_2$/Si ||||||||  | 300 nm SiO$_2$/Si ||||||
|---|---|---|---|---|---|---|---|---|---|---|---|---|---|---|
|  | $C_D$ || $C_{DR}$ || $C_{DG}$ || $C_{DB}$ || $C_D$ || $C_{DR}$ || $C_{DG}$ ||
|  | Mean value | SD | Mean value | SD | Mean value | SD | Mean value | SD | Mean value | SD | Mean value | SD | Mean value | SD |
| 1L | -10.4 | 0.5 | -9.0 | 1.4 | -12.8 | 1.1 | -8.8 | 2.4 | -3.7 | 0.8 | -2.9 | 0.7 | -8.3 | 0.8 |
| 2L | -22.5 | 0.5 | -20.0 | 1.7 | -27.4 | 0.9 | -19.4 | 3.8 | -6.6 | 0.5 | -6.0 | 0.8 | -17.2 | 0.5 |
| 3L | -34.3 | 0.8 | -29.1 | 1.0 | -42.3 | 1.2 | -32.7 | 2.6 | -9.8 | 0.4 | -6.9 | 0.8 | -24.3 | 1.0 |
| 4L | -46.3 | 0.7 | -38.2 | 2.8 | -59.6 | 1.8 | -41.1 | 3.4 | -13.3 | 0.9 | -10.0 | 1.4 | -31.4 | 1.6 |
| 5L | -57.2 | 0.7 | -47.8 | 4.4 | -75.6 | 2.2 | -48.9 | 9.0 | -15.9 | 1.0 | -15.3 | 1.0 | -36.2 | 1.1 |
| 6L | -67.9 | 0.8 | -55.4 | 2.3 | -91.0 | 1.1 | -57.1 | 5.8 | -18.6 | 0.5 | -19.2 | 1.4 | -38.7 | 0.9 |
| 7L | -77.2 | 0.4 | -64.9 | 2.4 | -105.6 | 1.1 | -62.8 | 4.4 | -21.5 | 0.7 | -25.3 | 1.7 | -41.8 | 1.2 |
| 8L | -86.4 | 0.5 | -77.5 | 3.6 | -119.4 | 0.8 | -63.7 | 5.9 | -24.7 | 1.0 | -31.9 | 2.7 | -44.4 | 0.4 |
| 9L | -93.2 | 1.5 | -88.6 | 1.6 | -130.8 | 1.3 | -63.1 | 4.4 | -27.1 | 1.0 | -35.4 | 1.1 | -46.9 | 0.4 |
| 10L | -100.6 | 0.4 | -95.6 | 0.9 | -141.2 | 0.7 | -67.3 | 1.5 | -28.5 | 0.7 | -38.9 | 0.9 | -47.4 | 1.4 |
| 11L | -107.6 | 0.5 | -106.6 | 4.2 | -153.1 | 1.9 | -66.8 | 1.8 | -31.3 | 0.3 | -43.3 | 0.7 | -50.4 | 0.5 |
| 12L | -111.1 | 1.0 | -116.8 | 0.9 | -156.1 | 1.3 | -65.8 | 1.6 | -31.7 | 0 | -46.6 | 0 | -50.4 | 0 |
| 13L | -116.0 | 0.7 | -118.8 | 0 | -160.0 | 0 | -67.5 | 0 | -33.5 | 0.8 | -48.6 | 1.5 | -53.0 | 0.9 |
| 14L | -117.3 | 0.3 | -127 | 0 | -157.7 | 0 | -65.8 | 0 |  |  |  |  |  |  |
| 15L | -119.7 | 0 | -136.8 | 0 | -158.3 | 0 | -64.7 | 0 |  |  |  |  |  |  |

**Table S2.** The optical contrast difference ($C_D$, $C_{DR}$, $C_{DG}$ and $C_{DB}$) values of MoS$_2$ nanosheets with different layer numbers on 90 and 300 nm SiO$_2$/Si. Optical images of MoS$_2$ nanosheets on 90 and 300 nm SiO$_2$/Si were taken at the exposure time of 80 and 50 ms, respectively.

| | 90 nm SiO$_2$/Si | | | | | | | | 300 nm SiO$_2$/Si | | | | | | | |
|---|---|---|---|---|---|---|---|---|---|---|---|---|---|---|---|---|
| | $C_D$ | | $C_{DR}$ | | $C_{DG}$ | | $C_{DB}$ | | $C_D$ | | $C_{DR}$ | | $C_{DG}$ | | $C_{DB}$ | |
| | Mean value | SD | Mean value | SD | Mean value | SD | Mean value | SD | Mean value | SD | Mean value | SD | Mean value | SD | Mean value | SD |
| 1L | -36.3 | 0.8 | -47.9 | 2.9 | -38.7 | 2.4 | -28.1 | 2.0 | -8.9 | 0.2 | -28.8 | 1.8 | -9.3 | 1.5 | 12.1 | 1.0 |
| 2L | -53.6 | 0.5 | -93.3 | 2.3 | -55.4 | 2.6 | -19.9 | 1.3 | -17.9 | 0.2 | -55.9 | 0.8 | -15.7 | 2.2 | 19.7 | 0.9 |
| 3L | -52.9 | 0.6 | -123.4 | 0.8 | -47.0 | 1.3 | 3.4 | 1.4 | -21.8 | 0.3 | -88.1 | 0.5 | -7.9 | 1.6 | 30.5 | 0.7 |
| 4L | -41.1 | 0.3 | -137.9 | 2.2 | -21.8 | 0.5 | 29.5 | 0.9 | -21.7 | 0.4 | -111.3 | 0.9 | 6.0 | 1.3 | 40.1 | 0.6 |
| 5L | -21.8 | 0.5 | -125.0 | 3.7 | 5.4 | 0.9 | 54.1 | 0.8 | -17.0 | 0.5 | -124.0 | 0.4 | 28.1 | 2.1 | 49.0 | 0.6 |
| 6L | 1.0 | 0.7 | -110.2 | 0.3 | 31.8 | 1.0 | 73.8 | 1.9 | -8.8 | 0.2 | -139.9 | 0.6 | 45.7 | 2.5 | 58.9 | 1.1 |
| 7L | 23.0 | 0.9 | -77.1 | 2.7 | 53.3 | 0.7 | 89.9 | 1.2 | -4.4 | 0.2 | -136.1 | 1.3 | 62.3 | 3.5 | 64.0 | 0.9 |
| 8L | 44.3 | 0.3 | -46.3 | 2.0 | 71.5 | 1.3 | 101.0 | 1.5 | 4.7 | 0.3 | -121.1 | 2.0 | 72.2 | 0.7 | 66.5 | 0.5 |
| 9L | 61.4 | 0.7 | -13.0 | 0.6 | 85.9 | 0.6 | 110.7 | 1.2 | 16.0 | 0.4 | -108.4 | 0.3 | 85.3 | 1.3 | 70.1 | 0.8 |
| 10L | 73.8 | 0.7 | 10.1 | 0.6 | 93.8 | 0.8 | 116.5 | 2.0 | 27.0 | 0.6 | -89.8 | 0.7 | 96.8 | 0.9 | 76.4 | 0.3 |
| 11L | 85.7 | 0.7 | 32.6 | 1.0 | 101.4 | 1.0 | 125.6 | 1.0 | 35.9 | 0.8 | -74.6 | 0.4 | 98.2 | 3.3 | 78.2 | 2.4 |
| 12L | 95.4 | 0.5 | 51.0 | 0.8 | 107.5 | 1.0 | 131.9 | 0.3 | 44.3 | 1.9 | -52.5 | 0.9 | 110.0 | 2.0 | 79.2 | 0.9 |
| 13L | 104.5 | 0.8 | 61.6 | 1.2 | 109.8 | 1.0 | 137.8 | 0.9 | 53.4 | 0.6 | -35.9 | 2.1 | 115.9 | 0.5 | 81.5 | 3.1 |
| 14L | 109.0 | 0.3 | 68.2 | 0.7 | 112.3 | 0.1 | 140.6 | 0.1 | 60.7 | 0.7 | -17.4 | 5.8 | 119.9 | 0.7 | 81.6 | 5.0 |
| 15L | 111.80063 | 0.1 | 77.0 | 0.3 | 122.3 | 0.1 | 136.8 | 3.7 | 68.8 | 0 | -1.3 | 0 | 121.5 | 0 | 82.1 | 0 |

**Table S3.** The optical contrast difference ($C_D$, $C_{DR}$, $C_{DG}$ and $C_{DB}$) values of WSe$_2$ nanosheets with different layer numbers on 90 and 300 nm SiO$_2$/Si. Optical images of WSe$_2$ nanosheets on 90 and 300 nm SiO$_2$/Si were taken at the exposure time of 80 and 50 ms, respectively.

|     | 90 nm SiO$_2$/Si ||||||||| 300 nm SiO$_2$/Si |||||||||
|     | $C_D$ || $C_{DR}$ || $C_{DG}$ || $C_{DB}$ || $C_D$ || $C_{DR}$ || $C_{DG}$ || $C_{DB}$ ||
|     | Mean value | SD | Mean value | SD | Mean value | SD | Mean value | SD | Mean value | SD | Mean value | SD | Mean value | SD | Mean value | SD |
|---|---|---|---|---|---|---|---|---|---|---|---|---|---|---|---|---|
| 1L  | -38.4 | 1.7 | -43.0  | 2.1 | -46.3 | 2.7 | -27.6 | 1.4 | -13.7 | 0.6 | -41.9  | 2.4 | -9.3  | 1.3 | 11.0329  | 1.1 |
| 2L  | -59.2 | 1.4 | -80.0  | 0.9 | -76.8 | 1.4 | -21.6 | 0.5 | -25.2 | 0.7 | -69.5  | 2.4 | -28.8 | 2.6 | 17.1861  | 1.0 |
| 3L  | -62.9 | 1.5 | -103.0 | 0.4 | -76.5 | 1.9 | -13.9 | 1.2 | -31.5 | 0.6 | -100.7 | 1.2 | -21.2 | 2.3 | 23.55704 | 1.4 |
| 4L  | -55.3 | 1.4 | -107.9 | 1.4 | -62.7 | 1.9 | 0.3   | 0.8 | -31.8 | 0.2 | -119.7 | 2.8 | -12.9 | 2.6 | 30.29348 | 0.8 |
| 5L  | -36.8 | 2.4 | -93.6  | 1.2 | -34.7 | 0.7 | 16.7  | 0.9 | -26.0 | 0.6 | -132.0 | 2.2 | 18.6  | 2.6 | 35.01708 | 0.4 |
| 6L  | -14.2 | 2.1 | -71.3  | 1.8 | -5.5  | 0.6 | 32.4  | 1.5 | -18.6 | 0.6 | -128.8 | 1.3 | 32.8  | 1.2 | 41.65274 | 1.6 |
| 7L  | 7.8   | 1.7 | -37.0  | 1.3 | 14.1  | 1.3 | 48.1  | 1.3 | -9.2  | 0.9 | -117.0 | 2.1 | 47.7  | 1.5 | 44.15263 | 1.1 |
| 8L  | 28.6  | 0.9 | -9.2   | 1.5 | 36.6  | 0.4 | 53.2  | 2.7 | 0.8   | 0.2 | -102.3 | 1.7 | 55.3  | 2.5 | 49.98872 | 1.8 |
| 9L  | 46.4  | 2.1 | 22.3   | 2.0 | 61.2  | 0.9 | 59.1  | 0.5 | 13.4  | 1.2 | -81.7  | 3.1 | 68.0  | 2.2 | 53.38465 | 2.2 |
| 10L | 61.8  | 1.5 | 48.4   | 1.3 | 71.7  | 0.9 | 66.2  | 0.3 | 22.7  | 0.7 | -52.9  | 2.9 | 77.6  | 2.9 | 54.51789 | 2.1 |
| 11L | 74.5  | 0.9 | 64.4   | 1.0 | 82.7  | 0.9 | 77.4  | 1.5 | 33.9  | 0.5 | -33.9  | 2.6 | 82.5  | 3.0 | 61.04936 | 5.7 |
| 12L | 82.4  | 0.9 | 73.6   | 1.3 | 87.7  | 1.1 | 85.8  | 1.1 | 43.3  | 0.1 | -27.7  | 2.9 | 89.3  | 2.7 | 64.37087 | 1.4 |
| 13L | 89.3  | 1.0 | 79.6   | 1.5 | 96.5  | 1.3 | 92.2  | 0.5 | 51.3  | 1.8 | -11.4  | 0   | 90.7  | 0   | 66.02392 | 0   |
| 14L | 95.5  | 0   | 82.9   | 0   | 101.2 | 0   | 101.0 | 0   | 55.3  | 0   | 2.4    | 0   | 91.7  | 0   | 66.27921 | 0   |

**Table S4.** The optical contrast difference ($C_D$, $C_{DR}$, $C_{DG}$ and $C_{DB}$) values of TaS$_2$ nanosheets with different layer numbers on 90 nm SiO$_2$/Si. Optical images were taken at the exposure time of 80 ms.

|  | 90 nm SiO$_2$/Si | | | | | | | |
|---|---|---|---|---|---|---|---|---|
|  | $C_D$ | | $C_{DR}$ | | $C_{DG}$ | | $C_{DB}$ | |
|  | Mean value | SD | Mean value | SD | Mean value | SD | Mean value | SD |
| 2L | -3.6 | 2.2 | -13.5 | 0.1 | -6.0 | 0.3 | 11.9 | 0.2 |
| 3L | -12.1 | 1.8 | -18.6 | 0.2 | -16.2 | 1.5 | 0.4 | 1.6 |
| 4L | -18.5 | 3.0 | -36.1 | 2.8 | -24.8 | 5.1 | -4.5 | 3.7 |
| 5L | -26.2 | 0 | -39.9 | 0 | -34.5 | 0 | -7.7 | 0 |
| 6L | -43.3 | 0.7 | -54.1 | 0.7 | -56.0 | 0.4 | -18.9 | 0.6 |
| 7L | -55.7 | 1.2 | -67.4 | 1.1 | -71.0 | 0.1 | -27.1 | 0.2 |
| 8L | -63.3 | 0 | -74.9 | 0 | -82.8 | 0 | -34.7 | 0 |
| 9L | -72.9 | 0.1 | -81.1 | 0.5 | -92.3 | 0.5 | -45.1 | 0.6 |
| 10L | -73.2 | 1.4 | -77.6 | 2.1 | -93.2 | 1.8 | -49.6 | 2.0 |
| 11L | -75.7 | 0.5 | -75.5 | 2.5 | -94.2 | 1.6 | -55.6 | 0.7 |
| 12L | -75.0 | 0 | -71.8 | 0 | -92.7 | 0 | -60.0 | 0 |
| 13L | -72.8 | 1.0 | -66.5 | 0.6 | -89.2 | 1.2 | -63.3 | 1.8 |
| 14L | -68.4 | 0 | -59.3 | 0 | -82.9 | 0 | -65.1 | 0 |
| 15L | -63.6 | 1.1 | -50.4 | 1.5 | -74.8 | 1.1 | -65.7 | 0.8 |
| 16L | -58.4 | 1.1 | -41.6 | 0.9 | -67.1 | 1.2 | -66.4 | 1.9 |
| 17L | -52.7 | 0 | -33.8 | 0 | -58.5 | 0 | -64.4 | 0 |
| 18L | -49.3 | 0.2 | -26.4 | 0.6 | -52.9 | 1.5 | -67.8 | 0.5 |
| 19L | -42.0 | 2.5 | -17.6 | 2.4 | -42.0 | 2.6 | -64.4 | 2.1 |
| 20L | -36.7 | 1.1 | -10.5 | 0.7 | -34.6 | 1.5 | -65.0 | 0.5 |
| 21L | -30.5 | 0.7 | -2.3 | 1.2 | -26.3 | 1.1 | -62.1 | 1.2 |
| 22L | -24.5 | 0.9 | 7.3 | 2.7 | -17.6 | 3.7 | -58.8 | 2.8 |
| 23L | -20.0 | 0 | 12.5 | 0 | -12.0 | 0 | -57.7 | 0 |
| 24L | -13.1 | 1.4 | 22.3 | 0.8 | -5.2 | 1.2 | -56.2 | 1.7 |
| 25L | -5.3 | 0 | 32.2 | 0 | 5.0 | 0 | -50.3 | 0 |
| 26L | -1.9 | 1.9 | 36.3 | 2.4 | 9.7 | 2.7 | -49.4 | 2.4 |
| 27L | 3.3 | 0.4 | 41.9 | 0.7 | 15.5 | 1.3 | -46.1 | 2.3 |

| 28L | 8.8 | 0.4 | 48.7 | 1.0 | 22.2 | 0.3 | -44.0 | 1.2 |
| 32L | 33.8 | 0 | 73.3 | 0 | 51.4 | 0 | -24.9 | 0 |

## 11. Thickness of various 2D nanosheets measured by AFM.
**Table S5.** Thicknesses of graphene, MoS$_2$ and WSe$_2$ nanosheets with different layer numbers measured by AFM.

|  | Graphene | | MoS$_2$ | | WSe$_2$ | | TaS$_2$ | |
|---|---|---|---|---|---|---|---|---|
|  | Mean value (nm) | SD | Mean value (nm) | SD | Mean value (nm) | SD | Mean value (nm) | SD |
| 1L | 0.4 | 0.1 | 0.7 | 0.1 | 0.7 | 0.1 | | |
| 2L | 0.7 | 0.1 | 1.4 | 0.1 | 1.5 | 0.1 | 1.4 | 0.1 |
| 3L | 1.0 | 0.1 | 2.1 | 0.1 | 2.1 | 0.1 | 1.9 | 0.1 |
| 4L | 1.4 | 0.1 | 2.7 | 0.1 | 2.8 | 0.1 | 2.6 | 0.1 |
| 5L | 1.7 | 0.1 | 3.5 | 0.1 | 3.5 | 0.1 | 3.3 | 0.1 |
| 6L | 2.0 | 0.1 | 4.0 | 0.1 | 4.1 | 0.1 | 4.0 | 0.1 |
| 7L | 2.4 | 0.1 | 4.6 | 0.1 | 4.7 | 0.1 | 4.6 | 0.1 |
| 8L | 2.8 | 0.1 | 5.2 | 0.1 | 5.4 | 0.1 | 5.3 | 0.1 |
| 9L | 3.1 | 0.1 | 5.9 | 0.1 | 6.0 | 0.1 | 6.0 | 0.1 |
| 10L | 3.4 | 0.1 | 6.5 | 0.1 | 6.7 | 0.1 | 6.6 | 0.1 |
| 11L | 3.7 | 0.1 | 7.1 | 0.1 | 7.4 | 0.2 | 7.2 | 0.1 |
| 12L | 4.1 | 0.1 | 8.0 | 0.1 | 8.1 | 0.1 | 8.0 | 0.1 |
| 13L | 4.4 | 0.1 | 8.7 | 0.1 | 8.8 | 0.1 | 8.5 | 0.1 |
| 14L | 4.8 | 0.1 | 9.3 | 0.1 | 9.4 | 0.1 | 9.1 | 0.1 |
| 15L | 5.2 | 0.1 | 10.0 | 0.1 | | | 9.8 | 0.1 |
| 16L | | | | | | | 10.5 | 0.1 |
| 17L | | | | | | | 10.9 | 0.1 |
| 18L | | | | | | | 11.6 | 0.1 |
| 19L | | | | | | | 12.3 | 0.1 |
| 20L | | | | | | | 12.9 | 0.1 |
| 21L | | | | | | | 13.6 | 0.1 |
| 22L | | | | | | | 14.2 | 0.1 |
| 23L | | | | | | | 14.9 | 0.1 |
| 24L | | | | | | | 15.5 | 0.1 |
| 25L | | | | | | | 16.1 | 0.1 |
| 26L | | | | | | | 16.7 | 0.1 |
| 27L | | | | | | | 17.5 | 0.1 |
| 28L | | | | | | | 18.1 | 0.1 |

| 32L | | | | | | 20.7 | 0.2 |

**12. The thickness of graphene nanosheets with minimum positive optical contrast difference on 90 and 300 nm SiO$_2$/Si.**

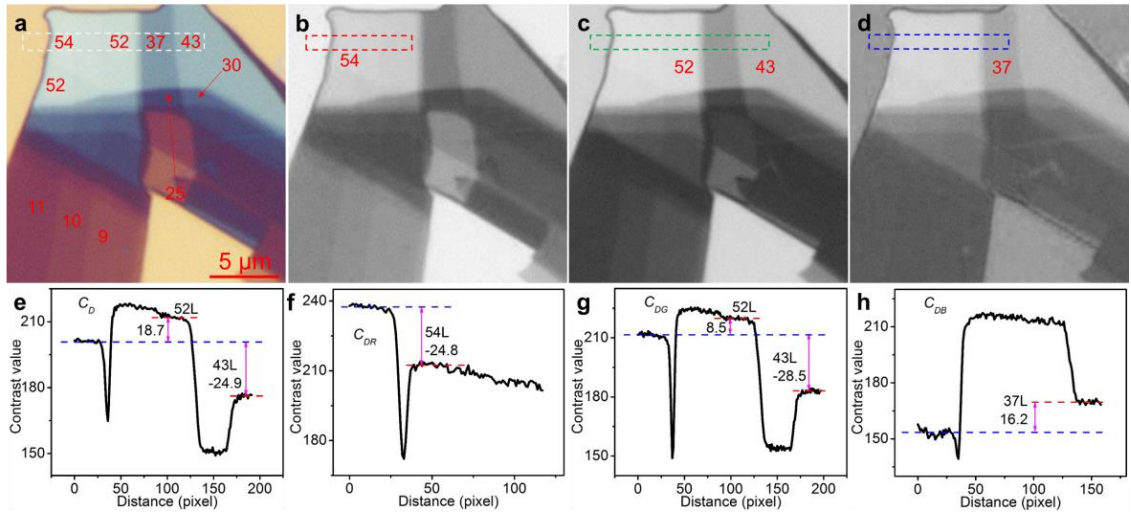

**Figure S12.** Color optical images (a) and grayscale images of R (b), G (c), and B (d) channels of graphene nanosheets on 90 nm SiO$_2$/Si and the corresponding contrast profiles (e-h) of the dashed rectangles shown in (a-d), respectively. The digitals shown in (a-d) indicate the layer numbers of corresponding graphene nanosheets.

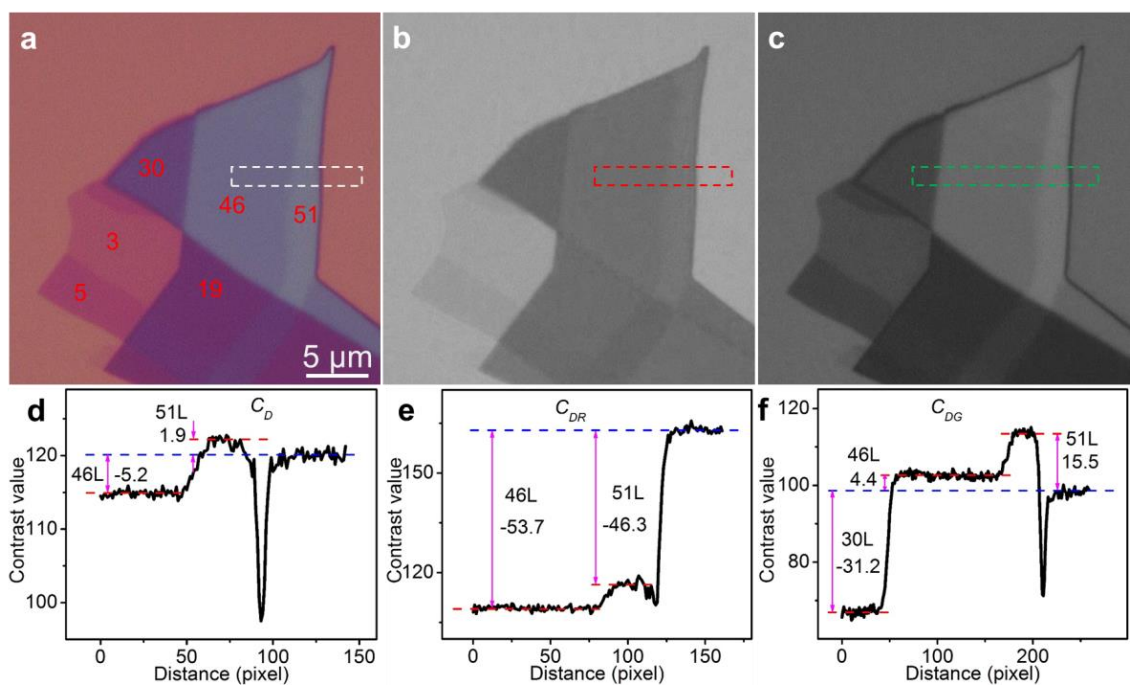

**Figure S13.** Color optical images (a) and grayscale images of R (b) and G (c) channels of graphene nanosheets on 300 nm $SiO_2$/Si and the corresponding contrast profiles (d-f) of the dashed rectangles shown in (a-c), respectively. The digitals shown in (a) indicate the layer numbers of corresponding graphene nanosheets.

49